\documentclass[aps,prl,showpacs,draft,twocolumn]{revtex4}
\usepackage{amsmath,bm}

\begin{document}

\title{Decay of scalar turbulence revisited}

\author{M. Chertkov$^a$ and V. Lebedev$^{a,b}$}

\affiliation{$^a$ Theoretical Division, LANL, Los Alamos, NM 87545, USA \\
$^b$ Landau Institute for Theoretical Physics, Moscow, Kosygina 2,
117334, Russia}


\date{\today}

\begin{abstract}

We demonstrate that at long times the rate of passive scalar decay
in a turbulent, or simply chaotic, flow is dominated by regions
(in real space or in inverse space) where mixing is less
efficient. We examine two situations. The first is of a spatially
homogeneous stationary turbulent flow with both viscous and
inertial scales present. It is shown that at large times scalar
fluctuations decay algebraically in time at all spatial scales
(particularly in the viscous range, where the velocity is smooth).
The second example explains chaotic stationary flow in a
disk/pipe. The boundary region of the flow controls the long-time
decay, which is algebraic at some transient times, but becomes
exponential, with the decay rate dependent on the scalar diffusion
coefficient, at longer times.

\end{abstract}

\pacs{47.27.Qb}

\maketitle

Study of advection of a steadily supplied scalar in smooth flow,
pioneered by Batchelor \cite{59Bat} for the case of a slowly
changing strain and extended by Kraichnan to the opposite limit of
short-correlated flow \cite{67Kra}, has developed into a universal
theory, applicable to any kind of statistics and temporal
correlations of the smooth chaotic flow \cite{94SS,95CFKLa}. (See
also the reviews \cite{00SS,01FGV}.) Extension of this theory to
the decay problem \cite{99Son,99BF} suggested a fast (exponential)
decay of the passive scalar in the case of spatially smooth (that
is approximated by linear velocity profiles at relevant scales)
and unbounded random flow. (We will call this case the ``pure
Batchelor" one.) In parallel, laboratory experiments
\cite{96MD,97WMG,00JCT,01GSb} and numerical simulations
\cite{93Pie,94HS,95AFO,97BDY} were conducted to verify the
Batchelor-Kraichnan theory. However, a comparison between theory,
experiment and numerical simulations was not conclusive. Different
experiments and simulations showed different results, inconsistent
with each other and with the theory. In this letter we explain a
possible source of the discrepancy. Our major point is that the
Batchelor-Kraichnan theory  does not apply to the late-time stage
of the scalar decay measured in the decay laboratory and numerical
experiments \cite{93Pie,96MD,97WMG,01GSb}. This is because the
long-time decay is dominated by the regions of the flow, where the
mixing is not as efficient as in the pure Batchelor case.

Before constructing a quantitative theory, let us draw a
qualitative physical picture. We start from the first example of a
multi-scale turbulent flow with the following hierarchy of scales
assumed: $L_v\gg\eta\gg r_d$, where $L_v$ is the energy containing
scale of the flow, $\eta$ is the Kolmogorov (viscous) scale and
$r_d$ is the diffusive scale of the scalar. It is known that the
passive scalar mixing in the inertial range of turbulent flow is
slower at larger scales, since the turnover time grows with scale.
This explains the algebraic-in-time decay of the scalar for the
inertial range of scales, $L_v\gg r\gg\eta$ \cite{00EX,01CEFV}. If
we now take into account that turbulent velocity is smooth at
scales smaller than $\eta$, it becomes important to understand
what happens with the scalar fluctuations at these small scales.
We show in the letter that scalar decay is accompanied by both
upscale and downscale transport, however, the situation is
asymmetric. The decay of scalar fluctuations in time is due to
inertial interval physics, while the spatial correlations of the
scalar at the smallest scales are mainly controlled by the viscous
part of the flow. (Here, the spatial correlations of the scalar
are logarithmic, like in the stationary case \cite{95CFKLa}.) An
analogous situation is realized if the scalar is advected by a
chaotic (spatially smooth) flow in a box or a pipe. Advection of
the scalar slows down in the boundary region, while mixing
continues to be efficient in the bulk, i.e. far from the
boundaries. The result is that the boundary region supplies the
scalar into the bulk. Plumes (blobs) of the scalar, injected from
the boundary domain into the box center, cascade in a fast way
down to $r_d$ where diffusion smears it out. Therefore, the
temporal decay of the scalar is mainly controlled by the rate of
scalar injection from the boundary layer. This causes an essential
slow down of the decay in comparison with the pure Batchelor case,
since in this case there are no stagnation (boundary) regions. The
decay of the scalar is algebraic during the transient stage, when
the scalar is supplied to the bulk from the boundary region which
is wider than the diffusive boundary layer. The boundary region
width decreases with time. Whenever the boundary layer width
becomes of the order of the diffusive layer width the decrease
stops and the algebraic temporal decay turns into an exponential
one. The exponential decay rate appears to be parametrically
smaller than in the pure Batchelor case.

In all the cases discussed, we consider decay of a passive scalar,
$\theta$, described by the equation
 \begin{eqnarray}
 \partial_t\theta+({\bm v}\nabla)\theta
 =\kappa\nabla^2\theta,
 \label{theta} \end{eqnarray}
where $\bm v$ is the flow velocity and $\kappa$ is the diffusion
coefficient. Our goal is, starting from the equation (\ref{theta})
and assuming some statistical properties of the velocity field $\bm v$,
to describe statistics of the passive scalar decay. In order to
understand the qualitative features of the passive scalar decay,
we choose (in the spirit of Kraichnan \cite{94Kra}) the simplest
possible case of a short-correlated in time velocity, possessing
Gaussian statistics. This model enables us to obtain analytically
a time evolution of the passive scalar correlation functions. Two
situations are discussed. The first is of a statistically homogeneous
flow (infinite volume) with a multi-scale velocity which is smooth
at small scales, and possesses a scaling behavior at larger scales.
The Reynolds number $Re=(L_v/\eta)^{4/3}$ and
the Prandtl number $Pr=(\eta/r_d)^2$ are both assumed to be large.
The second situation is $2d$ chaotic flow confined within a disk.
We also discuss an extension to the case of a more realistic pipe
flow. The flow is assumed to be smooth and obeying the standard
viscous behavior in the neighborhood of the boundary: the
longitudinal and transverse components of the velocities tend to
zero quadratically and linearly (respectively) with the distance
to the boundary.

We begin with an analysis of the homogeneous unbounded case. The
statistics of a short-correlated in time Gaussian velocity is
completely characterized by its pair correlation function
 \begin{eqnarray}
 \langle v_\alpha(t_1,{\bm r}) v_\beta(t_2,0)\rangle
 =\delta(t_1-t_2) \left[V_0\delta_{\alpha\beta}
 -{\cal K}_{\alpha\beta}(\bm r)\right],
 \label{BK1} \\
 {\cal K}_{\alpha\beta}\equiv \frac{r}{2}
 W'(r)\left(\delta_{\alpha\beta}
 -\frac{r_\alpha r_\beta}{r^2}\right)
 +\frac{d-1}{2}W(r)\delta_{\alpha\beta},
 \label{BK2} \end{eqnarray}
where $d$ is dimensionality of space, $V_0$ is a constant
characterizing fluctuations of the velocity at the integral
scale, $L_v$, and $W(r)$ is a function, which determines the
scale-dependence of velocity fluctuations. To describe both
viscous and inertial ranges, we introduce
 \begin{eqnarray}
 W=Dr^2/[1+(r/\eta)^\gamma] \,,
 \label{BK3} \end{eqnarray}
where $D$ stands for the amplitude of the velocity fluctuations
and $\gamma$ characterizes the velocity scaling in the inertial
interval. The Kolmogorov scale $\eta$ separates the inertial and
viscous intervals. The diffusive length is now expressed as
$r_d=\sqrt{\kappa/D}$. Measuring time, $t$, in the units of $1/D$,
one replaces $D$ by unity in all the forthcoming formulas. The
object we examine here is the simultaneous pair correlation
function of the passive scalar, $F(t,{\bm r})= \langle\theta(t,\bm
r)\theta(t,0)\rangle$, which is the simplest non-zero correlation
function in the homogeneous case. Under the assumption of isotropy
one derives from Eqs. (\ref{theta}-\ref{BK2})
 \begin{eqnarray}
 -r^{1-d}\partial_r\left\{r^{d-1}\left[W(r)+\kappa\right]
 \partial_r F_\lambda\right\}=\lambda F_\lambda,
 \label{Flambda} \end{eqnarray}
where $F_\lambda(r)$ is the Laplace transform of $F(t,r)$ with
respect to $t$. The operator on the left-hand side of Eq.
(\ref{Flambda}) is self-adjoint and non-negative (with respect to
the measure $\int\mbox dr\, r^{d-1}$ and under condition that
$F(r)$ is smooth at the origin and $\partial_r F(r=0)=0$).

We assume that $\int\mbox d\bm r\, \theta=0$ (the Corrsin invariant
is zero), and also, that initially the scalar is correlated at the
scale $r_0$ from the viscous-convective range, $r_d\ll r_0\ll \eta$,
i.e. $F(0,r)$ is $\approx F(0,0)$ at $r<r_0$ and decays fast enough
at $r>r_0$. Then some part of the initial evolution of $F(t,r)$ is
not distinguishable from the decay in the pure Batchelor case,
described by the limit $\eta\to\infty$ in Eq.(\ref{BK3}). Let us
briefly describe this initial stage of the dacay. The solutions of
Eq. (\ref{Flambda}) are power-like, $F_\lambda\propto
r^{\pm\sqrt{d^2/4-\lambda}-d/2}$, outside the diffusive range,
$r\gg r_d$. The functions are normalizable only if $\lambda>d^2/4$.
Therefore there is a gap in the spectrum
($0<\lambda<d^2/4$ are forbidden). The existence of this gap
guarantees an exponential decay of $F(t,r)$ with time. This feature
is, of course, in agreement with the previous analysis of the
pure Batchelor case \cite{99Son,99BF}. A complete set of functions
at $r\gg r_d$ is $F_\lambda=r^{-d/2}\cos[\alpha_k+k\ln(r/r_d)]$,
where $k$ is a positive number related to $\lambda$ via
$\lambda=d^2/4+k^2$, and $\alpha_k$ are phases, determined
by matching at $r\sim r_d$. Taking into account the orthogonality
condition for $F_\lambda$, $\int\mbox dr\,r^{d-1}F_\lambda(r)
F_{\lambda'}(r)=\pi\delta(k-k')/2$, one can express $F(t,r)$ via $F(0,r)$.
Analysis of this expression leads to the following picture. The
correlation function, $F(t,r)$, does not change at $r<r_-=r_0\exp(- td)$.
At the scales larger than $r_-$, $F$ decays exponentially,
$\propto\exp[-t d^2/4]$. Note, that $r_-$, also, marks the position of
the maximum of the spatial derivative of $F(t,r)$,
$\partial_r F(t,r)$. Therefore, $r_-(t)$ describes the
front running downscale (from $r_0$). The front reaches $r_d$ at
$t_d=\ln(r_0/r_d)/d$, so that at $t>t_d$, $F(t,r)$ decays
exponentially at all the small scales. Complimentary to the front running
downscale, there exists another front running upscale. Indeed, the
integral quantity, $\int_0^s\mbox dr\,r^{d-1}F(t,r)$, which determines
the overall amount of the scalar at all the scales smaller than
$s$, achieves its maximum around $r_+=r_0\exp(td)$.

At $t>t_\eta\sim\ln[\eta/r_0]/D$ the pure Batchelor description ceases
to be valid and one does need to account for complete multiscale form
of eddy-diffusivity function $W(r)$ given by Eq. (\ref{BK3}).
Our further analysis is devoted to this general case. The spatial
decay of the functions $F_\lambda$ in the multi-scale case is not
as steep as in its pure Batchelor one. It is convenient
to change to a new field, $\Phi_\lambda$,
$F_\lambda=r^{1-d}\partial_r[r^{d/2} \Phi_\lambda(z)]$,
where $z=2\sqrt{\lambda}r^{\gamma/2}/\gamma$. The general solution
of the resulting equation for $\Phi_\lambda(z)$ is a linear combination
of the Bessel functions, $\Phi_\lambda= J_{\pm\nu}(z)$, where
$\nu=2\sqrt{d^2/4 -\lambda}/\gamma$. If $\lambda<d^2/4$, the
positive root for $\Phi_\lambda$ has to be chosen to satisfy the
matching conditions at $r_d$. The eigen function is
normalizable for any positive $\lambda$, so that even the smallest
positive $\lambda$ are not forbidden. The lack of the gap in the
spectrum of the multi-scale model means an algebraic in time
decay of $F(t,r)$, i.e. much slower decay than the one found
for the Batchelor-Kraichnan model. We present here the expression
for the long-time, $t\gg t_\eta$, asymptotic behavior of $F(t,r)$:
 \begin{eqnarray}
 F(t,r)\!\propto \!\left\{
 \begin{array}{cc}
 \left[t d+\ln(r/\eta)
 \!+\!(r/\eta)^\gamma/\gamma\right]^{-1-d/\gamma},
 & r\ll r_+; \\
 t^{-5/4-d/2\gamma}r^{-d/2+\gamma/4}, & r\gg r_+
 \end{array}\right.
 \label{F2multi} \end{eqnarray}
Here $r_+=(\gamma t)^{1/\gamma}$ stands for position of the front
running upscale, $r_+$ lies in the scaling region $r_+\gg \eta$,
where the multi-scale model turns into the so-called Kraichnan
model \cite{94Kra}, $W\to r^{2-\gamma}$. (Therefore, it is not
surprising, that our result (\ref{F2multi}) is consistent
at $r\gtrsim r_+$ with the expression for the pair correlation
function derived before for the Kraichnan model \cite{00EX,01CEFV}
in the regime of zero Corrsin invariant.) Note, that when
$t-t_\eta$ is yet moderate in value, the small-scale part of the
asymptotic expression (\ref{F2multi}) is correct only at
$r\gg R_-(t)=\eta e^{-td}$, where $R_-$ is the position of the
front initiated at $t\sim t_\eta$ and running downscale from $\eta$.
At $t\sim t_\eta+\ln(\eta/r_d)/d$, the downscale front reaches the
dissipative scale, and afterwards the expression (\ref{F2multi}) is
correct for any scales $r\gg r_d$. Eq. (\ref{F2multi}) shows
that just like in the pumped case \cite{59Bat,67Kra,95CFKLa} the
scalar structure function is logarithmic, $\langle[\theta(t,\bm r)
-\theta(t,0)]^2\rangle \sim t^{-2-d/\gamma}\ln[r/r_d]$, at
$r_d\ll r\ll \eta$ and large times, where, therefore, the
time-dependent factor at the logarithm can be interpreted as a
scalar flux from the inertial range, $r\gtrsim \eta$, downscale
to the viscous range, $r\lesssim \eta$.

To conclude, the existence of the inertial region in velocity affects
drastically the spatio-temporal distribution of the scalar in the viscous range
(where the velocity can be approximated by linear profiles). The inertial
interval, where the scalar is mainly concentrated, serves as a kind of reservoir
for the smaller scales. This explains why the decay of scalar correlation
function in the viscous is much slower (algebraic) than in the pure Batchelor
case (when it would be exponential).

Now we proceed to the situation of spatially bounded flows. Aiming to
demonstrate the main qualitative features of the confined geometry, that is
strong sensitivity of the scalar decay to the peripheral (close to the boundary)
part of the flow, we examine a model case of $2d$ chaotic (i.e. consisting of
only few spatial harmonics) flow inside a disk. (It is, actually, clear that the
picture of temporal decay and spatial distribution of the scalar, described
below, is universal, i.e. it applies to a typical chaotic flow confined in a
close box, particularly in $3d$.) Incompressible flow in $2d$ can be
characterized by a stream function $\psi$, then the radial and azimutal
components of the velocity are $v_r=-\partial_\varphi\psi/r$ and $v_\varphi
=\partial_r\psi$. Our model of $\psi$ is
 \begin{eqnarray}
 \psi=-\frac{\xi_1}{2}r^2 U(r)\sin(2\varphi)
 +\frac{\xi_2}{2}r^2U(r)\cos(2\varphi) \,,
 \label{streamU} \\
 \langle\xi_i(t_1)\xi_j(t_2)\rangle
 =2D\delta_{ij}\delta(t_1-t_2) \,,
 \label{xi} \end{eqnarray}
where $U(r)$ is a function of $r$, finite at the origin, $U(0)=1$,
and becoming zero, together with its first derivative, at the disk
boundary, $r=1$, and $\xi_{1,2}(t)$ are zero mean short correlated random
Gaussian functions. The value of the passive scalar, $\theta$, averaged over the
statistics of $\xi$, $\langle\theta(t,\bm r)\rangle$, is the object of our
interest here. One examines how the average concentration of the scalar
evolves with time at different locations ${\bf r}$ within the disk.
The short-correlated feature of the velocity field allows a closed
description for $\langle\theta\rangle$ in terms of a partial
differential equation. Considering the spherically
symmetric part of $\langle\theta\rangle$ only (asymptotically,
at large times only this $varphi$-independent part remains essential), and
passing from $\langle\theta\rangle$ and $r$ to $\Upsilon$ and $q$,
respectively, where $\langle\theta\rangle= r^{-1}\partial_r
[r^2\Upsilon(t,q)]$ and $q=-\ln r$, one finds that the Laplace
transform of $\Upsilon$ with respect to $t$ (measured, again,
in $D^{-1}$ units) satisfies
 \begin{eqnarray} &&
 (U^2+\kappa e^{-2q})
 (\partial_q^2\Upsilon_\lambda -2\partial_q\Upsilon_\lambda)
 +\lambda\Upsilon_\lambda=0.
 \label{Philambda} \end{eqnarray}
The diffusion term is important only at the center of the disk and
near the boundaries. Analysis of Eq. (\ref{Philambda}) is straightforward
but bulky. Below we will present only selected details of the analysis,
aiming to describe the general picture of the phenomenon. (The complete
account for the derivation details will be published elsewhere \cite{02CLT}.)

In the bounded flow the decay of passive scalar splits into three distinct
stages. The major effect dominating the first stage (just as in the pure
Batchelor case, explained by Eq. (\ref{Philambda}) with $U\to1$) is formation of
elongated structures (stripes) of the scalar in the bulk region of the flow. The
stripes are getting thinner with time, i.e. inhomogeneities of smaller and
smaller scales are produced. Once the width of the stripes decreases down to the
dissipative scale, $r_d\sim\sqrt{\kappa}$, the stripes are smeared out by
diffusion. The stretching-contraction process is exponential in time, so that
the initial stage when the stripes are formed lasts for $\tau_1\sim\ln[1/r_d]$.
By the end of this first stage the scalar is exhausted in the central region of
the flow. The stretching rate, however, is smaller in the peripheral domain than
in the bulk. Thus the scalar literally remains longer in the peripheral domain.
This defines the second, transient, stage. At $\kappa^{1/4}\ll q\approx 1-r
\ll\lambda^{1/4}\ll 1$, the solution of Eq. (\ref{Philambda}) is
$\Upsilon_\lambda\propto q \sin\left(\alpha_\lambda +\sqrt\lambda q/U\right)$.
In the other asympotitc region at $q\gg\lambda^{1/4}$ one can drop the second
derivative (with respect to $q$) and diffusive terms in Eq. (\ref{Philambda}).
One finds that, $\partial_q\ln\Upsilon_\lambda\sim 1$ at $q\sim\lambda^{1/4}$.
Matching those two asymptotic regions at $q\sim \lambda^{1/4}$ we come to the
$\alpha_\lambda=0$ condition. Once the asymptotic form of the eigen-function
$\Upsilon_\lambda$ is known, it is straightforward to restore the behavior of
$\Upsilon(t,q)$. One finds that $\Upsilon(t,q)$ (and, therefore,
$\langle\theta\rangle$) is concentrated in the $\delta(t)\sim 1/\sqrt{t}$-small
(and shrinking with time) vicinity of the boundary. In the domain outside of the
shrinking layer the decay is algebraic, $\langle\theta\rangle\propto
t^{-3/2}q^{-3}$. The second stage lasts for $\tau_2\sim\kappa^{-1/2}$, i.e.
until $\delta(t)$ shrinks down to $r_{bl}=\kappa^{1/4}$, which is the width of
the diffusive boundary layer. Account for diffusivity in the $q\ll 1$ analysis
gives yet another matching condition for $\Upsilon_\lambda$, at $r\sim r_{bl}$,
resulting in formation of a discrete spectrum for $\lambda$. The smallest
$\lambda$ (and therefore the value of the level spacing in the discrete
spectrum) is estimated by $\sqrt\kappa$. (Using the periodic character of the
$\sin$-function, one finds that the $n$-th level eigenvalue $\lambda_n$ is
estimated by $\sim n^2\sqrt\kappa$ at large $n$.) Therefore, at $t\gg\tau_2$,
$\langle\theta\rangle$ decays exponentially, $\propto\exp(-t/t_d)$, where
$t_d\sim\kappa^{-1/2}$. The smallest $\lambda$ eigen-function is localized at
$q\sim r_{bl}$. Thus the third (final) stage of evolution is characterized by
the majority of the scalar remaining in the diffusive boundary layer.

To conclude, the temporal behavior of the passive scalar
correlations, predicted by the bounded flow theory, is
complicated. It involves two different exponential regimes
separated by an algebraic one, so that a special accuracy is
required in order to quantify the theory against various
experiments in the field.

The bounded flow theory can be applied to the experiment of Groisman and Steinberg \cite{01GSb},
where the passive scalar is advected through a pipe by dilute-polymer-solution flow. The
polymer-related elastic instability makes the flow chaotic. The mean flow along the pipe is also
essential, so that according to the standard Taylor hypothesis, measurements of scalar
concentration at different positions along the pipe are interpreted as correspondent to different
times in an artificial decay problem. Exponential decay of the scalar was reported in \cite{01GSb}
at long times (long pipes). The theoretical picture explaining the experiment is similar to the
one proposed above for the disk. The width of the diffusive boundary layer near the pipe
boundaries is estimated by $R\cdot{Pe}^{-1/4}$, where $R$ is the pipe radius and $Pe$ is the
Peclet number, $Pe=R^2 \sigma/\kappa$ and $\sigma$ is a typical value of the velocity gradient
(${Pe}$ is $\sim 10^4$ in the conditions of \cite{01GSb}). The characteristic decay time (fixed by
the diffusive boundary layer width) is estimated by $t_d\sim\sigma^{-1}{Pe}^{1/2}$. However, one
should be careful when converting this time to the decrement of the passive scalar decay along the
direction of the flow (pipe). The average velocity of the flow tends to zero near the boundary of
the pipe, making the advection near the boundary less efficient than in the bulk. Assuming a
linear profile of the average velocity near the boundary, one gets the following estimate for the
law of scalar fluctuation decay along the pipe, $\sim\exp(-\gamma z/u_0)$, where $z$ is the
coordinate in the direction of the pipe, $u_0$ is the average velocity at the center of the pipe,
and $\gamma\sim\sigma\,{Pe}^{-1/4}$. The factor $Pe^{-1/4}$ here is a manifestation of a slow down
of the passive scalar decay in comparison with the pure Batchelor case. The same factor
characterizes decay of higher order correlation functions of the scalar as well. For example, one
can consider the passive scalar pair correlation function $F(r)$ near the center of the disk. One
finds that $F(r)\propto r^{-\alpha}$, where $\alpha$ is small and can be estimated as
$\alpha\sim\,\mbox{Pe}^{-1/4}$. (Note, that this behavior is almost indistinguishable from the
logarithmic one, found for the stationary case \cite{59Bat,67Kra,95CFKLa}.) This estimations agree
with the experimental data of \cite{01GSb}.

Let us now briefly discuss two other available numerical and
experimental results. Pierrehumbert \cite{93Pie} reported
exponential decay of the scalar correlations in his numerical
experiment with chaotic map velocity in a periodic box. This
observation is consistent with our results, since the periodic
boundary conditions are less restrictive than the zero condition
for velocity at the box boundary, and, therefore, should leave a
finite gap in the $\lambda$-spectrum even in the limit
$\kappa\to0$. In the experiment of Jullien, Castiglione and
Tabeling on $2d$ stationary flow steered by magnets in a finite
volume beneath the $2d$ layer \cite{00JCT}, the decay rate of the
scalar seems slower than exponential (Fig. 2 of \cite{00JCT}).
This observation is in agreement with the absence of the gap in
the $\lambda$-spectrum (at $\kappa\to0$) we found for the finite
box model.

The brevity of this letter does not allow us to discuss effects of
intermittency (which manifests in high-order moments of the
scalar), of higher order angular harmonics (in the disk or pipe
geometry), and details of the spatial distribution of the scalar
in a variety of other inhomogeneous cases (e.g. periodic flow).
This detailed discussion is postponed for a longer paper to be
published elsewhere \cite{02CLT} (where we will also present
results of direct numerical simulations of the passive scalar
decay for these cases).

We thank B. Daniel, R. Ecke, G. Eyink, G. Falkovich, A. Fouxon, A.
Groisman, I. Kolokolov, M. Riviera, V. Steinberg, P. Tabeling and
Z. Toroczkai for helpful discussions.

\end{document}